\documentclass[runningheads]{llncs}
\usepackage[T1]{fontenc}
\usepackage{graphicx}
\usepackage{booktabs} 
\usepackage{multirow}
\usepackage{amsmath}
\usepackage{multibib}
\usepackage{algorithm}
\usepackage{algpseudocode}
\usepackage{hyperref}
\usepackage[table,xcdraw]{xcolor}
\usepackage{multirow}
\usepackage{tabularx} 
\usepackage{xcolor}
\usepackage[table]{xcolor}
\usepackage{float}

\usepackage{parskip}

\algnewcommand\algorithmicforeach{\textbf{for each}}
\algdef{S}[FOR]{ForEach}[1]{\algorithmicforeach\ #1\ \algorithmicdo}
\begin{document}
\title{Automatic Aortic Valve Pathology Detection from 3-Chamber Cine MRI with Spatio-Temporal Attention Maps}
\author{Y. On \inst{1} \and 
K. Vimalesvaran \inst{1} \and 
C. Galazis \inst{1} \and 
S. Zaman \inst{2} \and 
J. Howard \inst{1} \and 
N. Linton \inst{1} \and 
N. Peters \inst{1} \and 
G. Cole \inst{2} \and 
A.A. Bharath \inst{1} \and 
M. Varela \inst{1}}

\authorrunning{On et al.}
\titlerunning{Automatic Detection of Aortic Valve Pathology}

\institute{Imperial College London, Exhibition Road, London, SW7 2AZ, UK \and Imperial College Healthcare NHS Trust, Du Cane Road, London, W12 0HS, UK
\email{yu.on16@imperial.ac.uk}}
\date{January 2023}
\maketitle
\begin{abstract}
The assessment of aortic valve pathology using magnetic resonance imaging (MRI) typically relies on blood velocity estimates acquired using phase contrast (PC) MRI. However, abnormalities in blood flow through the aortic valve often manifest by the dephasing of blood signal in gated balanced steady-state free precession (bSSFP) scans (Cine MRI). We propose a $3$D classification neural network (NN) to automatically identify aortic valve pathology (aortic regurgitation, aortic stenosis, mixed valve disease) from Cine MR images. We train and test our approach on a retrospective clinical dataset from three UK hospitals, using single-slice $3$-chamber cine MRI from N = $576$ patients. Our classification model accurately predicts the presence of aortic valve pathology (AVD) with an accuracy of $0.85 \pm 0.03$ and can also correctly discriminate the type of AVD pathology (accuracy: $0.75 \pm 0.03$). Gradient-weighted class activation mapping (Grad-CAM) confirms that the blood pool voxels close to the aortic root contribute the most to the classification. Our approach can be used to improve the diagnosis of AVD and optimise clinical CMR protocols for accurate and efficient AVD detection. 
\end{abstract}
\newpage
\section{Introduction}
\subsubsection{Aortic Valve Disease}
Aortic valve disease (AVD) is one of the most common cardiovascular pathologies \cite{2022}. It affects ~4.5 million people in Europe, with a prevalence of 0.18\% \cite{10.1093/eurheartj/ehab892,10.1016/S0195-668X(03)00201-X}. The most common manifestations of AVD \cite{10.1093/eurheartj/ehab395,10.1007/978-3-031-16431-6_54} are:
\begin{itemize}
    \item aortic stenosis (AS) - a narrowing of the aortic valve, leading to a reduced aortic valve cross-sectional area and increased blood velocity during systole;
    \item aortic regurgitation (AR) - a leaking of the aortic valve, causing the blood to flow backwards into the left ventricle during diastole;
    \item mixed valve disease (MVD) - simultaneous presentation of AR and AS.
\end{itemize}
The assessment of AVD using MRI typically relies on blood velocity estimates acquired using 2D cardiac gated phase contrast (PC) MRI \cite{troger2022}, which can provide maps of blood velocity perpendicular to the aortic valve.

\vspace{-0.4cm}
\subsubsection{Cine MRI}
Cine MRI is arguably the most commonly used cardiac magnetic resonance (CMR) sequence, as it provides valuable information about cardiac function, anatomy and blood flow. Cine MRI usually employs a 2D bSSFP (balanced steady state free precession) readout, with thick slices ($8-10~mm$ thick), each acquired in one breathhold and is typically gated to image the cardiac cycle at 20 or more cardiac phases \cite{doi:10.2214/ajr.148.2.239}. Single slices are commonly acquired in 4-chamber, 2-chamber and 3-chamber (also known as left ventricular outflow) orientations, as well as stacks of short axis slices for full ventricular coverage \cite{2ffcaffec0fa47c4930d841e33feeaeb}. 
 
Blood is typically bright in bSSFP images given its comparatively high $T_2/T_1$ ratio and bSSFP's relative insensitiveness to flow due to the inherent first-order velocity compensation of most of its gradients. Turbulent flow or high blood speeds can however lead to the loss of phase coherence between consecutive radio-frequency pulses, creating dark regions in the blood pool \cite{https://doi.org/10.1002/mrm.20619}. These flow void regions can be exploited to help diagnose blood flow pathology \cite{https://doi.org/10.1002/jmri.23544}, but to date this assessment has relied on subjective visual assessment or, more recently, machine learning tools reliant on expert manual annotations\cite{10.1007/978-3-031-16431-6_54}.

\vspace{-0.4cm}
\subsubsection{Aim} We propose a deep learning approach to automatically detect aortic valve pathology (AS, AR, and MVD) from 3-chamber Cine CMR, using multi-binary labels without the need for any expert annotation of imaging features (such as aortic valve insertion points).

\section{Methods}
\subsubsection{Imaging Data}
We train and evaluate our method on a retrospective clinical CMR dataset, obtained from different scanner manufacturers across three hospitals in the UK. We use $N=576$ single-slice bSSFP 3-chamber Cine MR images. The resolution of the images is $1.17-1.56 \times 1.17-1.56 \time 8~ mm^3$ and the number of frames is $22 \pm 10$ (range: 12-32). Each image is labelled into four different classes using information from clinical records.  In total, $387$ patients ($67$\%) have no aortic valve abnormalities ("no pathology"). Of the remaining $189$, $83$ ($14\%$) have AR, $56$ ($10\%$) AS, and $50$ ($9\%$) MVD. We train a 3D convolutional neural network (CNN) for two separate classification tasks: 2-class (labels: "no pathology" or "AVD") and 4-class classification (labels: "no pathology", "AR", "AS" or "MVD").

\begin{figure}[!h]
\centering
\includegraphics[width=1.0\textwidth]{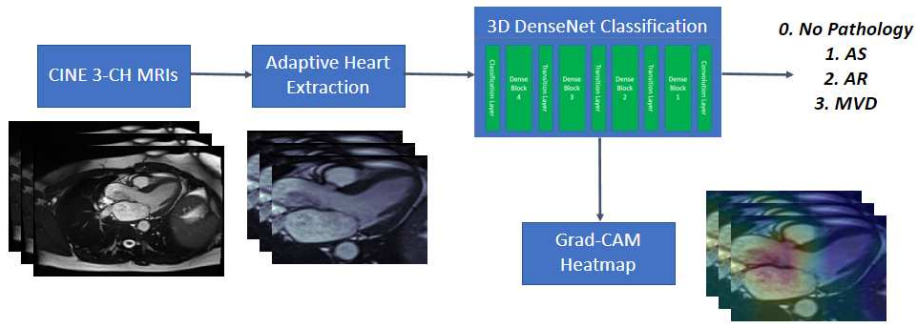}
\caption{Overview of our automatic method for aortic valve pathology detection from cine MRI. Each of the steps is described in the text in detail.
}\label{fig:overview}
\end{figure}

Our pipeline consists of three main steps, as shown in Fig. \ref{fig:overview}: (i) adaptive heart extraction in all $3-chamber$ Cine MRI frames, (ii) image classification into either 2 or 4 classes, and (iii) identification the voxels that contributed the most to the classification using Grad-CAM (Gradient-weighted Class Activation Mapping) heatmaps \cite{DBLP:journals/corr/SelvarajuDVCPB16}.

\vspace{-0.3cm}

\vspace{-0.4cm}
\subsubsection{Adaptive Heart Extraction}
\begin{figure}[!h]
\centering
\includegraphics[width=1.0\textwidth]{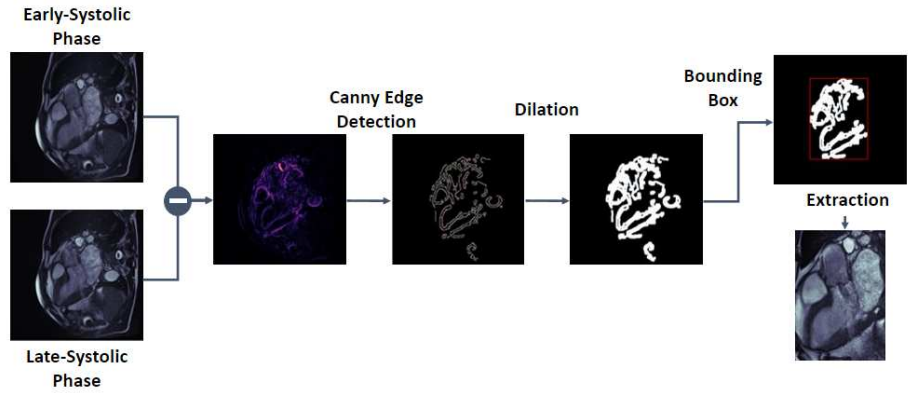}
\caption{The adaptive heart detection algorithm. The early-systolic phase is subtracted from the late-systolic phase, before undergoing Canny edge detection and morphological dilation. The largest connected component in this image is extracted and used to create a bounding box around the heart.
}\label{fig:extraction}
\end{figure}

We propose a heuristic algorithm (see Fig. \ref{fig:extraction}) to automatically identify the cardiac structure in all Cine MRI frames, based on the assumption that the heart is the largest image structure that moves across the cardiac cycle. First, the algorithm calculates the absolute difference of the early systolic and late systolic cardiac phases. A Canny edge detector ($\sigma = 2.0$) \cite{4767851} is then applied to identify the edges separating the different structures in the difference image, which undergo morphological dilation with a $2$-pixel radius diamond structuring element. The largest connected component in the dilated edge image is identified as the heart and a tight bounding box is drawn around it. To standardise the images for preparation to the classification model, two post-processing steps are applied. The extracted region is re-sampled to a spatial resolution of $1 \times 1$ mm, the images are padded with zeros or cropped further for a consistent image size of $224 \times 224 \times D$ ($D$: original number of frames). 

\vspace{-0.3cm}

\vspace{-0.3cm}

\subsubsection{Classification Neural Network} 
\begin{figure}[!h]
\centering
\includegraphics[width=1.0\textwidth]{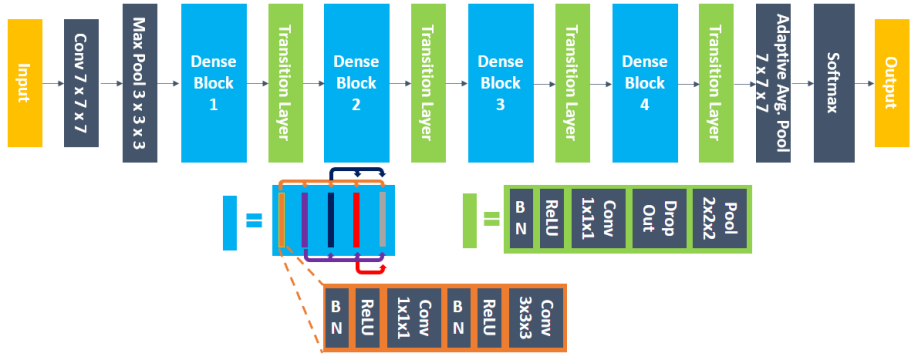}
\caption{A diagram of the 3D DenseNet architecture used. It consists of 4 identical dense blocks separated by transitional layers. Each dense block includes 5 consecutive operations, each including $1 \times 1 \times 1$ and $3 \times 3 \times 3$ convolutional layers ("Conv"). Within the dense block, the feature maps from each convolutional layer are concatenated to the output of the next layers. The transition layers downsample the feature maps produced by the dense blocks and include batch normalisation ("BN"), a $1 \times 1 \times 1$ convolutional layer, a $2 \times 2 \times 2$ average pooling layer and is placed between two consecutive dense blocks to facilitate the downsampling operation. Rectified linear units ("ReLU") are used throughout as activation layers. The feature vector is then flattened and classified into one of the classes with a softmax function.}\label{fig:densenet}
\end{figure}

A 3D DenseNet model \cite{DBLP:journals/corr/HuangLW16a} based on the prototype available in the MONAI library \cite{https://doi.org/10.48550/arxiv.2211.02701} is used to classify the 2D+time Cine MR images. As shown in Fig. \ref{fig:densenet}, the model has 4 identical dense blocks separated by transitional layers. The dense blocks extract image features with a given size by concatenating the outputs of five successive convolutional operations. They are separated by transitional layers which combine and downsample the features outputted by each dense block before passing them to the next one. After the final dense block, an adaptive average pooling layer is implemented with a n-class softmax layer to perform the classification (see Fig. \ref{fig:densenet}).

\vspace{-0.3cm}

\vspace{-0.3cm}

Prior to the classification, we perform z-score normalisation \cite{DBLP:journals/corr/PatroS15} and 'on-the-fly' data augmentation. We use data augmentation techniques such as rotation, contrast and bias field adjustment \cite{SUDRE201750} to minimise overfitting \cite{https://doi.org/10.48550/arxiv.2211.02701}. Each augmentation is individually applied with a probability of 0.2. In order to cope with the variability in the number of frames, the input is area interpolated \cite{CARUSO1998109} to a depth of 30 to synchronise all the input dimensions to $224 \times 224 \times 30$.

\vspace{-0.4cm}
\subsubsection{Experimental Setting and Performance Criteria}
We train the CNN on 1 GeForce RTX6000 graphical processing units (NVIDIA, Santa Clara, California) for approximately 11 hours. We optimise our CNN using the Adam (learning rate: $0.0001$) and train it for $500$ epochs with a batch size of $2$. Focal loss \cite{DBLP:journals/corr/abs-1708-02002} function is used in both 2- and 4-class classification. The focal loss function is expected to be more robust for the imbalanced labels present in the classification tasks. We assess the classification performance of our method in the testing data using: accuracy, F$_1$ score, precision and Area Under the Receiver Operating Characteristics Curve (ROC AUC) \cite{doi:10.1177/0272989X8400400203}. The 576 data is randomly split into training (322), validation (81), and test (173) sets in a stratify fashion based on the class labels.

\vspace{-0.4cm}
\subsubsection{Grad-CAM}
After the classification, we use Grad-CAM heatmaps \cite{DBLP:journals/corr/SelvarajuDVCPB16,DBLP:journals/corr/abs-2007-00453} to visualise the gradients of a targeted class at a specific convolutional layer. Heatmaps are usually calculated using the gradients from a deep layer because of the high-level semantic information it encodes without trading off the spatial information. Also, because of the class discriminant nature of Grad-CAM, only the gradients from the targeted class are used while the gradients from other classes are set to zero. We use the last convolutional layer of the final transitional layer (see Fig. \ref{fig:densenet}) to generate the heatmap. An open-source library \cite{2007.00453} is utilised. The returned heatmap is normalised to a range between 0 (blue) and 1 (red), and rescaled to $224 \times 224 \times 30$ to match the classifier input size. For each label, the regions that contribute the most to the assigned class will be shown in red and the regions with less relevant contributions, in blue.

\section{Results}
\begin{figure}[H]
\centering
\includegraphics[width=0.8\textwidth]{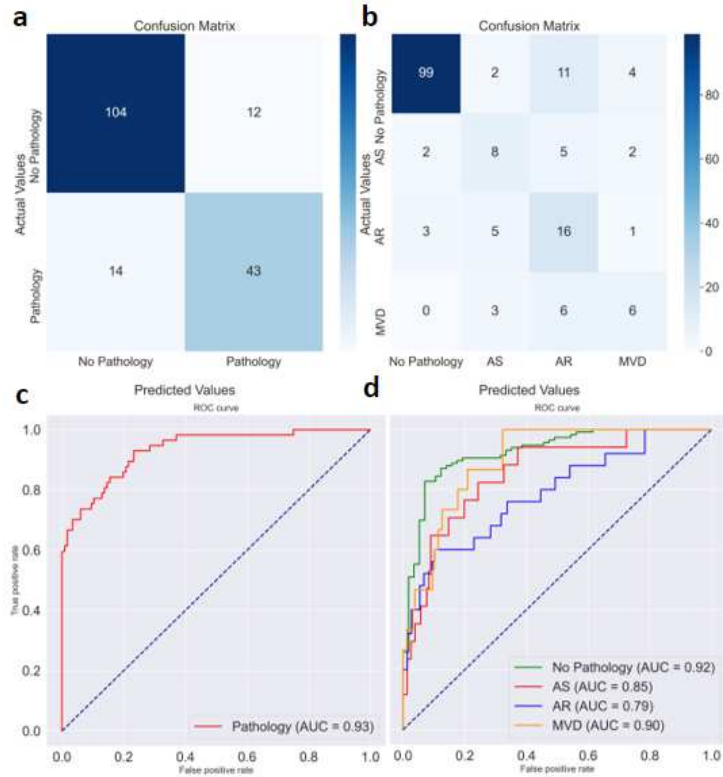}
\caption{The confusion matrix and ROC AUC of the classification results. (a, c) Confusion matrix and ROC AUC of the 2-Class classification (b, d) Confusion matrix and ROC AUC of the 4-Class classification. The x-axis of the confusion matrix shows the predicted labels versus the y-axis as the ground truth labels. The ROC in the multiclass classification is one versus all: the given class is treated as positive class with the rest treated as negative class.
}\label{fig:cm_n_roc}
\end{figure}

\begin{figure}[H]
\centering
\includegraphics[width=0.7\textwidth]{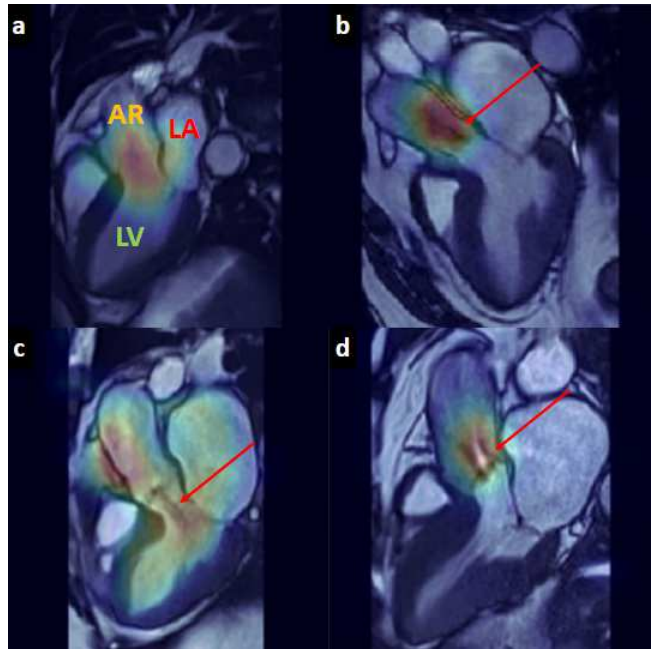}
\caption{3-chamber Cine MRI examples with corresponding Grad-CAM heatmaps from 4 subjects at late systolic phase except early diastolic phase for AR. We label the aortic root (AR), left atrium (LA) and left ventricle (LV) in one of the panels. (a) no pathology; (b) aortic stenosis - the arrow shows a region of blood signal dephasing caused by the high-speed flow through the calcified aortic valve; (c) aortic regurgitation - the arrow points to blood signal dephasing caused by mixing with the backwards flow; (d) mixed valve disease.}
\label{fig:examples}
\end{figure}

\begin{figure}[!h]
\centering
\includegraphics[width=1.0\textwidth]{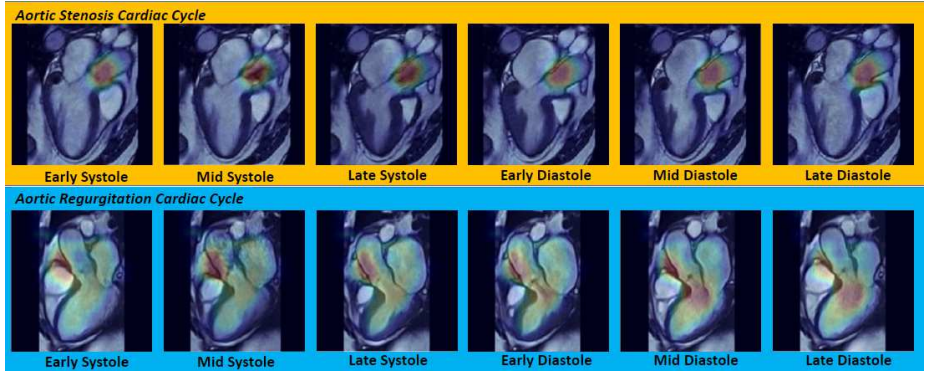}
\caption{Examples of 3-chamber Cine MR images across the cardiac cycle, overlaid with corresponding Grad-CAM heatmaps from 2 subjects with AS (top) and AR (bottom). (Top - AS) The heatmaps show a strong contribution of the aortic root region throughout, especially during systole. (Bottom - AR) During the diastolic phases, the classification network focuses on the backwards flow from the aorta to the LV.}
\label{fig:cardiac_cycle}
\end{figure}

Our proposed method can identify AVD using 3-chamber Cine MRI automatically and reliably, as shown in Table \ref{table:classification} and Fig. \ref{fig:cm_n_roc}. It performs particularly well for the simpler 2-class discrimination task ("AVD" vs "no pathology") where its mean accuracy reaches 0.85, compared to 0.75 for the 4-class task. AR is the most difficult pathology to identify, as shown in Fig. \ref{fig:cm_n_roc}d).
Examples of Grad-CAM heatmaps from the 4 classes are illustrated in Fig. \ref{fig:examples}. The red region in the heatmaps shows the region contributed the most to the chosen label, which are mostly in the region of the aortic root and along the flow jet, as expected. In Fig. \ref{fig:examples}b and d, which corresponds to AS and MVD, the red region mostly highlights the high velocity blood flow and aortic valve; whereas in Fig. \ref{fig:examples}c, which correspond to AR, the red regions focus at the aortic root and the area that blood flows backward into the left ventricle. These regions are similar to the regions that clinicians use to visually inspect to assist diagnosis.
This is further illustrated in Fig. \ref{fig:cardiac_cycle}. In the AS cases, GradCAM focuses on the aortic root region, especially during systole. In AR, on the other hand, the GradCAM tracks the backwards flow of blood into the ventricle in diastole.

\vspace{-0.3cm}

\begin{table}[H]
\centering
\begin{tabular}{p{2cm}p{2cm}p{2cm}}
\hline
\rowcolor[HTML]{FFEBCD} 
\textbf{Metrics}   & \textbf{2-Class} & \textbf{4-Class} \\ \hline
\textbf{Accuracy}    & $0.85 \pm 0.03$    & $0.75 \pm 0.03$ \\ \hline
\textbf{F$_1$ Score}    & $0.83 \pm 0.03$    & $0.57 \pm 0.03$ \\ \hline
\textbf{Precision}    & $0.83 \pm 0.03$    & $0.57 \pm 0.03$ \\ \hline
\end{tabular}
\caption{Evaluation metrics for the 2-class and 4-class classification. The values shown are mean $\pm$ standard deviation.  \label{table:classification}}
\end{table}

\vspace{-0.3cm}

\vspace{-0.3cm}

\vspace{-0.3cm}

\section{Conclusions}
We present a robust pipeline that can identify aortic valve pathology automatically from 3-chamber Cine MRI, without the need for manual annotations of the aortic valve features and blood flow and delineation of the heart. The classification pipeline (with visual inputs provided by the Grad-CAM heatmaps) can be used a triage tool to identify patients with suspected AV pathology who may benefit further CMRI investigations such as PC-MRI. The proposed analysis tools are part of an effort to personalise clinical CMR protocols, making them less time consuming and costly.

Previous work has used machine learning methods to identify AVD from 3-chamber Cine MRI, but relied on expert manual annotations as inputs to the classification method \cite{10.1007/978-3-031-16431-6_54}. In particular, Vimalesvaran \textit{et al} propose a multi-level network to identify aortic valve landmarks, followed by a random forest model for AVD pathology binary classification. Their study reached an accuracy and F$_1$ score of $0.93 \pm 0.03$ and $0.91 \pm 0.04$ respectively, comparable to $0.85 \pm 0.03$ and $0.83 \pm 0.03$ we achieve respectively. Our approach, however, relies on the direct application of a CNN to the images, avoiding the need for the time-consuming manually-annotated image features as inputs.

Underlying the good performance of our method is the proposed adaptive heart extraction algorithm. This is able to identify the heart in Cine MRI in a computationally low-cost, automatic setup without any need for segmentation labels. Our offline experiments suggested the proposed algorithm is robust even for patients with low cardiac function. This step may not be necessary when training the classifier on much larger datasets using a network with a higher capacity. Other options consist in using a supervised heart identification method or unsupervised approaches using Grad-CAM \cite{DBLP:journals/corr/SelvarajuDVCPB16}. 

To mitigate the 'black-box' nature of applying NN in the clinical environment, we apply Grad-CAM after the classification process. For all the considered labels, as illustrated in Figs. \ref{fig:examples} and \ref{fig:cardiac_cycle}, the model consistently relies on voxels from the blood pool in the aortic root and, for AR, also in the LV. This is reassuring as AVD patients often present with morphological changes in the heart \cite{thubrikar2018aortic}, which the CNN could have relied upon for its classification. Moreover, GradCAM (see Fig. \ref{fig:cardiac_cycle}) suggests that the classifier may also perform well when receiving a smaller number of cardiac phases as inputs, which could reduce the computational burden of the network.

Once trained, our approach can be used in real time at the scanner side, to identify patients with likely AVD who would benefit from additional dedicated imaging, such as aortic root PC-MRI scans. It can therefore contribute to the personalisation and optimisation of CMR protocols, leading to improved more efficient patient diagnosis. Future work will include the testing of the proposed tools on larger datasets from other sites, a comparison of our approach with different state-of-the-art CNNs, and classification of valvular pathology using Cine MRIs acquired in different views.

\textbf{Acknowledegments.} This work was supported by the British Heart Foundation Centre of Research Excellence at Imperial College London (RE/18/4/34215). K. Vimalesvaran and S. Zaman were supported by UKRI Centre for Doctoral Training in AI for Healthcare grant number EP/S023283/1. The project was partly supported by a Rosetrees Interdisciplinary Award.

\bibliographystyle{IEEEtran}

\bibliography{mybibfile}

\begin{thebibliography}{10}
\providecommand{\url}[1]{#1}
\csname url@samestyle\endcsname
\providecommand{\newblock}{\relax}
\providecommand{\bibinfo}[2]{#2}
\providecommand{\BIBentrySTDinterwordspacing}{\spaceskip=0pt\relax}
\providecommand{\BIBentryALTinterwordstretchfactor}{4}
\providecommand{\BIBentryALTinterwordspacing}{\spaceskip=\fontdimen2\font plus
\BIBentryALTinterwordstretchfactor\fontdimen3\font minus
  \fontdimen4\font\relax}
\providecommand{\BIBforeignlanguage}[2]{{%
\expandafter\ifx\csname l@#1\endcsname\relax
\typeout{** WARNING: IEEEtran.bst: No hyphenation pattern has been}%
\typeout{** loaded for the language `#1'. Using the pattern for}%
\typeout{** the default language instead.}%
\else
\language=\csname l@#1\endcsname
\fi
#2}}
\providecommand{\BIBdecl}{\relax}
\BIBdecl

\bibitem{2022}
\BIBentryALTinterwordspacing
M.~Guglielmo \emph{et~al.}, ``The role of cardiac magnetic resonance in aortic
  stenosis and regurgitation,'' \emph{Journal of Cardiovascular Development and
  Disease}, vol.~9, no.~4, p. 108, Apr 2022. [Online]. Available:
  \url{http://dx.doi.org/10.3390/jcdd9040108}
\BIBentrySTDinterwordspacing

\bibitem{10.1093/eurheartj/ehab892}
\BIBentryALTinterwordspacing
A.~Timmis \emph{et~al.}, ``{European Society of Cardiology: cardiovascular
  disease statistics 2021},'' \emph{European Heart Journal}, vol.~43, no.~8,
  pp. 716--799, 01 2022. [Online]. Available:
  \url{https://doi.org/10.1093/eurheartj/ehab892}
\BIBentrySTDinterwordspacing

\bibitem{10.1016/S0195-668X(03)00201-X}
\BIBentryALTinterwordspacing
B.~Iung \emph{et~al.}, ``{A prospective survey of patients with valvular heart
  disease in Europe: The Euro Heart Survey on Valvular Heart Disease},''
  \emph{European Heart Journal}, vol.~24, no.~13, pp. 1231--1243, 07 2003.
  [Online]. Available: \url{https://doi.org/10.1016/S0195-668X(03)00201-X}
\BIBentrySTDinterwordspacing

\bibitem{10.1093/eurheartj/ehab395}
\BIBentryALTinterwordspacing
A.~Vahanian \emph{et~al.}, ``{2021 ESC/EACTS Guidelines for the management of
  valvular heart disease: Developed by the Task Force for the management of
  valvular heart disease of the European Society of Cardiology (ESC) and the
  European Association for Cardio-Thoracic Surgery (EACTS)},'' \emph{European
  Heart Journal}, vol.~43, no.~7, pp. 561--632, 08 2021. [Online]. Available:
  \url{https://doi.org/10.1093/eurheartj/ehab395}
\BIBentrySTDinterwordspacing

\bibitem{10.1007/978-3-031-16431-6_54}
K.~Vimalesvaran \emph{et~al.}, ``Detecting aortic valve pathology
  from the 3-chamber cine cardiac mri view,'' in \emph{Medical Image
  Computing and Computer Assisted Intervention -- MICCAI 2022}.\hskip 1em plus
  0.5em minus 0.4em\relax Cham: Springer Nature Switzerland, 2022, pp.
  571--580.

\bibitem{troger2022}
F.~Troger \emph{et~al.}, ``A novel approach to determine aortic valve area with
  phase-contrast cardiovascular magnetic resonance,'' \emph{Journal of
  cardiovascular magnetic resonance : official journal of the Society for
  Cardiovascular Magnetic Resonance}, vol.~24, p.~7, 01 2022.

\bibitem{doi:10.2214/ajr.148.2.239}
\BIBentryALTinterwordspacing
U.~Sechtem \emph{et~al.}, ``Cine mr imaging: potential for the evaluation of
  cardiovascular function,'' \emph{American Journal of Roentgenology}, vol.
  148, no.~2, pp. 239--246, 1987, pMID: 3492096. [Online]. Available:
  \url{https://doi.org/10.2214/ajr.148.2.239}
\BIBentrySTDinterwordspacing

\bibitem{2ffcaffec0fa47c4930d841e33feeaeb}
C.~Kramer \emph{et~al.}, ``\BIBforeignlanguage{English}{Standardized
  cardiovascular magnetic resonance imaging (cmr) protocols: 2020 update},''
  \emph{\BIBforeignlanguage{English}{Journal of Cardiovascular Magnetic
  Resonance}}, vol.~22, Feb. 2020.

\bibitem{https://doi.org/10.1002/mrm.20619}
\BIBentryALTinterwordspacing
O.~Bieri and K.~Scheffler, ``Flow compensation in balanced ssfp sequences,''
  \emph{Magnetic Resonance in Medicine}, vol.~54, no.~4, pp. 901--907, 2005.
  [Online]. Available:
  \url{https://onlinelibrary.wiley.com/doi/abs/10.1002/mrm.20619}
\BIBentrySTDinterwordspacing

\bibitem{https://doi.org/10.1002/jmri.23544}
\BIBentryALTinterwordspacing
G.~Sommer, J.~Bremerich, and G.~Lund, ``Magnetic resonance imaging in valvular
  heart disease: Clinical application and current role for patient
  management,'' \emph{Journal of Magnetic Resonance Imaging}, vol.~35, no.~6,
  pp. 1241--1252, 2012. [Online]. Available:
  \url{https://onlinelibrary.wiley.com/doi/abs/10.1002/jmri.23544}
\BIBentrySTDinterwordspacing

\bibitem{DBLP:journals/corr/SelvarajuDVCPB16}
\BIBentryALTinterwordspacing
R.~R. Selvaraju \emph{et~al.}, ``Grad-cam: Why did you say that? visual
  explanations from deep networks via gradient-based localization,''
  \emph{CoRR}, vol. abs/1610.02391, 2016. [Online]. Available:
  \url{http://arxiv.org/abs/1610.02391}
\BIBentrySTDinterwordspacing

\bibitem{4767851}
J.~Canny, ``A computational approach to edge detection,'' \emph{IEEE
  Transactions on Pattern Analysis and Machine Intelligence}, vol. PAMI-8,
  no.~6, pp. 679--698, 1986.

\bibitem{DBLP:journals/corr/HuangLW16a}
\BIBentryALTinterwordspacing
G.~Huang, Z.~Liu, and K.~Q. Weinberger, ``Densely connected convolutional
  networks,'' \emph{CoRR}, vol. abs/1608.06993, 2016. [Online]. Available:
  \url{http://arxiv.org/abs/1608.06993}
\BIBentrySTDinterwordspacing

\bibitem{https://doi.org/10.48550/arxiv.2211.02701}
\BIBentryALTinterwordspacing
M.~J. Cardoso \emph{et~al.}, ``Monai: An open-source framework for deep
  learning in healthcare,'' 2022. [Online]. Available:
  \url{https://arxiv.org/abs/2211.02701}
\BIBentrySTDinterwordspacing

\bibitem{DBLP:journals/corr/PatroS15}
\BIBentryALTinterwordspacing
S.~G.~K. Patro and K.~K. Sahu, ``Normalization: {A} preprocessing stage,''
  \emph{CoRR}, vol. abs/1503.06462, 2015. [Online]. Available:
  \url{http://arxiv.org/abs/1503.06462}
\BIBentrySTDinterwordspacing

\bibitem{SUDRE201750}
\BIBentryALTinterwordspacing
C.~H. Sudre, M.~J. Cardoso, and S.~Ourselin, ``Longitudinal segmentation of
  age-related white matter hyperintensities,'' \emph{Medical Image Analysis},
  vol.~38, pp. 50--64, 2017. [Online]. Available:
  \url{https://www.sciencedirect.com/science/article/pii/S1361841517300257}
\BIBentrySTDinterwordspacing

\bibitem{CARUSO1998109}
\BIBentryALTinterwordspacing
C.~Caruso and F.~Quarta, ``Interpolation methods comparison,'' \emph{Computers
  and Mathematics with Applications}, vol.~35, no.~12, pp. 109--126, 1998.
  [Online]. Available:
  \url{https://www.sciencedirect.com/science/article/pii/S0898122198001011}
\BIBentrySTDinterwordspacing

\bibitem{DBLP:journals/corr/abs-1708-02002}
\BIBentryALTinterwordspacing
T.~Lin \emph{et~al.}, ``Focal loss for dense object detection,'' \emph{CoRR},
  vol. abs/1708.02002, 2017. [Online]. Available:
  \url{http://arxiv.org/abs/1708.02002}
\BIBentrySTDinterwordspacing

\bibitem{doi:10.1177/0272989X8400400203}
\BIBentryALTinterwordspacing
B.~J. McNeil and J.~A. Hanley, ``Statistical approaches to the analysis of
  receiver operating characteristic (roc) curves,'' \emph{Medical Decision
  Making}, vol.~4, no.~2, pp. 137--150, 1984, pMID: 6472062. [Online].
  Available: \url{https://doi.org/10.1177/0272989X8400400203}
\BIBentrySTDinterwordspacing

\bibitem{DBLP:journals/corr/abs-2007-00453}
\BIBentryALTinterwordspacing
K.~Gotkowski, C.~Gonz{\'{a}}lez, A.~Bucher, and A.~Mukhopadhyay, ``M3d-cam: {A}
  pytorch library to generate 3d data attention maps for medical deep
  learning,'' \emph{CoRR}, vol. abs/2007.00453, 2020. [Online]. Available:
  \url{https://arxiv.org/abs/2007.00453}
\BIBentrySTDinterwordspacing

\bibitem{2007.00453}
K.~Gotkowski \emph{et~al.}, ``M3d-cam: A pytorch library to generate 3d data
  attention maps for medical deep learning,'' 2020.

\bibitem{thubrikar2018aortic}
M.~Thubrikar, \emph{The aortic valve}.\hskip 1em plus 0.5em minus 0.4em\relax
  Routledge, 2018.

\end{thebibliography}

\end{document}